\documentstyle[fleqn] {revp}
\newfont{\sff}{cmssi12} % for affiliation and section italic
\newfont{\bigsf}{cmss12 scaled 2000} % for title
\newfont{\midsf}{cmss12 scaled 1000} % for title's subscription
\newfont{\smlsf}{cmss12 scaled 600}  
\newfont{\bigsff}{cmssi12 scaled 2000} % for title's italic
\newfont{\sfi}{cmssi10} % for section's italic

\newcommand{\pmonth}{August, 1995}
\newcommand{\spec}{Astrophysics}
\newcommand{\no}{10}

\begin{document}
\parindent 0pt
\parskip 12pt
\setcounter{page}{1}

\title{Compression modes in nuclei: 
%RPA and QRPA predictions with
microscopic models with 
Skyrme interactions}

\author{G. Col\`o$^{*1}$, N. Van Giai$^{*2}$, P.F. Bortignon$^{*1}$ and M.R.
Quaglia$^{*1}$ \\  
\sff
*1 Dipartimento di Fisica and INFN, via Celoria 16, I-20133 Milano (Italy)\\
\sff
*2 Groupe de Physique Th\'eorique, IPN, F-91406 Orsay (France)
}

\abst{The isoscalar giant monopole resonances (ISGMR) and giant dipole 
resonances 
(ISGDR) in medium-heavy nuclei are investigated in the framework of HF+RPA 
and 
HF-BCS+QRPA with Skyrme effective interactions. It is found that pairing has 
little 
effect on these modes. It is also found that the coupling of the RPA states 
to 2p-2h 
configurations results in about (or less than) 1 MeV shifts of the resonance 
energies 
and 
at the same time gives the correct total widths. 
For the ISGMR, comparison with recent data leads to a value of nuclear matter 
compression modulus close to 215 MeV. However, a discrepancy between 
calculated and measured energies of the ISGDR in $^{208}$Pb is found and 
remains an open 
problem.}

\maketitle
\thispagestyle{headings}

\section*{Introduction}

The determination of the nuclear incompressibility $K$ is still a matter of
debate, despite a remarkable number of works on the subject$^{1)}$. In the
present contribution, we present self-consistent calculations of the nuclear
collective modes associated with a compression and expansion of the nuclear
volume, namely the isoscalar giant monopole and dipole resonances (ISGMR and
ISGDR, respectively). In fact, we share the point of view that the most
reliable way to extract information on $K$ is to perform that kind of
calculations, having as the only phenomenological input a given effective
nucleon-nucleon interaction, and choose the value of $K$ corresponding to
the force which can reproduce the experimental properties of the compression
modes in finite nuclei.

The ISGMR, or ``breathing mode'', is excited by the operator
$\sum_{i=1}^A r_i^2$ and it has been identified in many isotopes along the 
chart
of nuclei already two decades ago. However, this systematics has never
allowed an unambigous determination of $K$$^{1)}$. This was one of the motivations
for the recent experimental program undertaken at the Texas A\&~M Cyclotron
Institute, which has allowed the extraction of experimental data for the
ISGMR of better quality as compared to the past, by means of the analysis
of the results of inelastic scattering of 240 MeV $\alpha$-particles. We
refer to other contributions in these proceedings for reports on these
experimental data$^{2)}$. Monopole strength functions turn
out to be quite fragmented for nuclei lighter than $^{90}$Zr. For nuclei 
like $^{208}$Pb, $^{144}$Sm, $^{116}$Sn and $^{90}$Zr, however, one is able
to identify a single peak which, together with a high-energy
extended tail, exhausts essentially all the monopole 
Energy Weighted Sum Rule (EWSR). These medium-heavy 
nuclei are, therefore, those suited for the extraction of information about
the nuclear incompressibility and we concentrate ourselves on them in the
present work.

The ISGDR is excited by the operator $\sum_i r_i^3 Y_{10}$ and corresponds
to a compression of the nucleus along a definite direction, so that it has
been called sometimes the ``squeezing mode''. Although some first indication
about the energy location of this resonance dates back to the beginning of
the eighties, a more clear indication about its strength distribution in
$^{208}$Pb has been reported only recently$^{3)}$. Measurements have been
done also for other nuclei, namely $^{90}$Zr, $^{116}$Sn and $^{144}$Sm (as
in the case of the giant monopole resonance). There is some expectation that
the study of this mode can help to shed some light on the problem of nuclear 
incompressibility. Actually, at first sight this compressional mode seems to 
provide us with a new problem. A  
simple assumption like the scaling model (illustrated for the present
purposes in Ref.$^{4)}$) would lead to two different values of the
finite-nucleus incompressibility $K_A$ if applied to the ISGMR and the ISGDR
with the input of their experimental energies. The hydrodynamical model 
gives two results which are closer$^{4)}$ but which still make us
wonder about the validity of methods based on extracting $K_A$ and
extrapolating it to large values of $A$, for the determination of $K$. This
points again to the necessity of reliable microscopic calculations of
the compressional modes, in order to reproduce the experimental data and 
extract the value of $K$ from the properties of the force which is used.  

Our calculations are performed within the framework of
self-consistent Hartree-Fock (HF) plus Random-Phase Approximation (RPA).
We use effective forces of Skyrme type$^{5-7)}$
and we look at their predictions for the properties of ISGMR and ISGDR. The 
parametrizations we employ span a large range of values for $K$ (from 200 MeV to
about 350 MeV). In particular, we focus mainly on two original
aspects: firstly, we look at the effects of pairing correlations in
open-shell nuclei; secondly, we study if the
picture obtained at mean-field level is altered by the inclusion of the
coupling of the giant resonances to more complicated nuclear configurations.
This inclusion 
%f their experimental energies. The hydrodynamical model again
is necessary if one wishes to understand theoretically all the
contributions to the resonance width and may shift the resonance centroid. 

About the first aspect, it is well known that pairing correlations are
important in general to explain the properties of ground states and 
%of open-shell nuclei. 
low-lying excited states in open-shell nuclei. 
Since we wish to see how these
correlations affect in particular the compressional modes, 
we take them into account by extending the HF-RPA
approach to a quasi-particle RPA (QRPA) on top of a HF-BCS calculation.

About the second aspect, 
%original issue mentioned, we briefly recall first that if
we recall that if 
we start from a description of the giant resonance as a superposition of
one particle-one hole (1p-1h) excitations, in their damping process we must
take care of the coupling with states of 2p-2h character. They are in fact
known to play a major role and give rise to the spreading width
$\Gamma^\downarrow$ of the giant resonance which is usually a quite 
large fraction of the total width. Within mean field theories, only the
width associated with the resonance fragmentation 
(Landau width) and the escape width
$\Gamma^\uparrow$ are included (the latter, provided that 1p-1h
configurations with the particle in the continuum are considered).
In the past, we have developed a theory in which all the contributions to the
total width of giant resonances are consistently treated and we have obtained
satisfactory results when applying it to a number of cases. In
particular, we will recall what has been obtained$^{8)}$ for the
case of the ISGMR in $^{208}$Pb. We
also report about a new calculation for the ISGDR in
the same nucleus.

%The paper is organized as follows. In sec. 2, we briefly survey our
%formalism. Since Skyrme HF+RPA is a widely used approach, described
%extensively in the literature, we just comment about the way we have
%extended it in the case of open-shell nuclei, and then we recall the basic
%structure of our microscopic theory of the damping process of giant
%resonances. In Sec. 3 and 4, we present the results for ISGMR and ISGDR,
%respectively. Finally, we draw some conclusions by pointing out the open
%questions, in particular those concerning the determination of the nuclear
%incompressibility.

\section*{Formalism: a brief survey}

For all nuclei we consider, we solve the HF equations on a radial mesh 
and, in the case of the open-shell isotopes, we solve HF-BCS equations. 
%In order to do that, 
A constant pairing gap $\Delta$ is introduced (for 
neutrons in the case of $^{116}$Sn and for protons in the case of 
$^{90}$Zr and $^{144}$Sm), and at each HF iteration the quasi-particle 
energies, the occupation factors and the densities to be input at the 
next iteration are determined accordingly. $\Delta$ is obtained from 
the binding energies of the neighboring nuclei$^{9)}$. The states 
included in the solution of the HF-BCS equations are those below a cutoff
energy given by $\lambda_{HF}+8.3$ MeV ($\lambda_{HF}$ being the HF Fermi
energy), in analogy with the procedure of Ref.$^{10)}$.

%Using these results for the nuclear ground-state, 
Using the above self-consistent mean fields 
we work out the RPA or 
QRPA equations (respectively on top of HF or HF-BCS), in their matrix form. 
Discrete positive energy states are obtained by diagonalizing the mean
field on a harmonic oscillator basis and they are used to build the 1p-1h (or
2 quasi-particles) basis coupled to $J^\pi$=0$^+$ or 1$^-$. The dimension 
of this basis is chosen in such a way that more than 95\% (typically 97-99\%)
of the appropriate EWSR is exhausted in the RPA or QRPA calculation. More 
details, especially on the way the QRPA equations are implemented, will be
given in Ref.$^{11)}$.

As mentioned in the previous section, in the case of $^{208}$Pb we perform
also calculations that go beyond this simple discrete RPA. This is done 
along the formalism described in Ref.$^{12)}$, which is recalled here only very 
briefly.

We label by $Q_1$ the space of discrete 1p-1h configurations in which the RPA 
equations are solved. To account for the escape width $\Gamma^\uparrow$ 
and spreading width $\Gamma^\downarrow$ of the giant resonances, we build two
other orthogonal subspaces $P$ and $Q_2$. The space $P$ is made of particle-hole 
configurations where the particle is in an unbound state orthogonal to 
all the discrete single-particle levels; the space $Q_2$ is built with the 
configurations which are known to play a major role in the damping process 
of giant resonances: these configurations are 1p-1h states coupled to a 
collective vibration. Using the projection operator formalism one can easily 
find that the effects of coupling the subspaces $P$ and $Q_2$ to $Q_1$ are 
described by the following effective Hamiltonian acting in the $Q_1$ space: 
\begin{eqnarray}
     {\cal H} (E) \equiv Q_1 H Q_1 & + & W^\uparrow(E)
     + W^\downarrow(E) \nonumber \\
     = Q_1 H Q_1 & + & Q_1 H P {\textstyle 1 \over \textstyle
     E - PHP + i\epsilon} P H Q_1 \nonumber \\
     \ & + & Q_1 H Q_2 {\textstyle 1 \over \textstyle
     E - Q_2 H Q_2 + i\epsilon}
     Q_2 H Q_1, \nonumber 
\label{H_eff}\end{eqnarray} 
where $E$ is the excitation energy. For each value of $E$ the RPA equations
corresponding to this effective, complex Hamiltonian ${\cal H} (E)$ are
solved and the resulting sets of eigenstates enable us to calculate all
relevant quantities, in particular the strength function associated with a
given operator. To evaluate the matrix elements of $W^\downarrow$,
we calculate the collective phonons with the same effective
interaction used for the giant resonance we are studying (within RPA),
and we couple these phonons with the 1p-1h components of the giant resonance
by using their energies and transition densities. 

\section*{Results for the isoscalar monopole resonance}

As recalled in the introduction, Youngblood {\it et al.}$^{2)}$ have
recently measured the ISGMR strength distribution with fairly good precision, 
in the nuclei $^{90}$Zr, $^{116}$Sn, $^{144}$Sm and $^{208}$Pb. In 
their work, they also compare the experimental centroid energies with the 
%outcome of 
calculations of Blaizot {\it et al.}$^{13)}$ performed 
by using RPA and employing the finite-range Gogny effective interaction: a
value of the nuclear incompressibility $K$ = 231 MeV is deduced. In the following, 
we denote as centroid energy the ratio $E_0\equiv m_1/m_0$ ($m_0$ and $m_1$ being 
the non-energy-weighted and energy-weighted sum rules, respectively). 

If we try to compare the experimental values with calculations done at the 
same RPA level but using the zero-range Skyrme effective interactions, we 
can infer a different conclusion with respect to the value of $K$. Among 
the Skyrme type interactions, the parametrization which gives probably the
best account of the experimental centroid energies in the nuclei studied
by the authors of Ref.$^{2)}$, is the SGII force$^{6)}$. 
%This can be seen in 
The results are shown in 
Table 1. The force SGII is characterized by a value of the nuclear 
incompressibility $K$ = 215 MeV: since it reproduces very well the ISGMR 
centroid energy in $^{208}$Pb, and it slightly overestimates those in the 
other isotopes, one would conclude that $K$ is of the order of or slightly
less than 215 MeV. 

This conclusion is inferred by means of simple RPA. It is of course legitimate
to wonder if calculations beyond this simple approximation could lead to 
different values of the nuclear incompressibility. We first consider the effect 
of pairing correlations. In the case of $^{116}$Sn, the centroid energy of
17.18 MeV obtained with the SGII force in RPA, becomes 17.19 MeV if one 
turns to QRPA. A very small shift is found also when other forces are
used (for instance, with the recently proposed 
SLy4 force$^{7)}$, one 
obtains 17.51 MeV and 17.59 MeV for RPA and QRPA, respectively) 
%must compare an RPA centroid energy of 17.51 MeV with a
%QRPA value of 17.59) 
and when other nuclei are considered. In general,
although we know that pairing correlations play a crucial role not only to
explain the ground-state of open-shell nuclei but also their low-lying excited 
states, it appears that they do not affect so much the giant resonances like
the ISGMR (or ISGDR, anticipating results of the next section) which
lie at relatively high excitation energy compared to the pairing gap $\Delta$.
Civitarese {\it et al.}$^{14)}$ found also small shifts (of the order
of 100-150 keV) for the ISGMR and ISGQR when pairing correlations are 
taken into account: this shift is larger than that obtained in the present
work, but it is the result of a different (non 
self-consistent) model. 

The present conclusion for the nuclear incompressibility is therefore similar 
to that obtained by Hamamoto {\it et al.}$^{15)}$, since they find that 
the Skyrme interaction which provides the best results for the ISGMR is the
SKM$^*$ parametrization and this is very similar to SGII (the associated 
nuclear incompressibility being 217 MeV). 
%But our conclusion lies on a somehow firmer ground 
Our study, however, is done in a more general framework 
since we have analyzed also the role of pairing correlations. 

If we finally consider the results of calculations beyond mean field$^{8)}$  
(which include not only the continuum coupling but also the coupling with the 
2p-2h type states) performed 
%by some of us in the case of 
for $^{208}$Pb we find that 
%in this case 
it is also possible to reproduce rather well the total width of the
ISGMR, which is around 3 MeV. This width is actually in 
large part a consequence of fragmentation (or Landau damping): at least three 
states share, at the level of RPA, the resonance strength, but continuum as 
well as 2p-2h couplings are able to give to each peak the correct width so that 
the overall lineshape coincides with the experimental findings. We stress that 
the coupling with the 2p-2h type states is also responsible
for a downward shift of the ISGMR centroid and peak energies, which is of the 
order of 0.5 MeV. One may argue that this affects the extraction of the
value of $K$ from theoretical calculations. Actually, since the value of $K$ 
associated with a given force is obtained by a calculation of nuclear matter 
at the mean field level, it is legitimate to draw conclusions about $K$ from
the comparison with the experiment of the 
ISGMR results for finite nuclei obtained again at the mean field level. But 
the fact that a given force is able to account for the ISGMR linewidth enforces
our confidence about its reliability. And it would be of course legitimate 
either, to compare the centroid energies obtained after 2p-2h coupling with 
experiment provided the value of $K$ associated with the force is calculated 
by including the same couplings at the nuclear matter level. No such calculations
in nuclear matter have been done so far, to our knowledge.  

\begin{table}[here]
\caption{Experimental and theoretical values of the centroid energies $E_0\equiv
m_1/m_0$ for the ISGMR and ISGDR. The theoretical values are obtained with 
the Skyrme-RPA approach, using the SGII interaction. 
%The discussion in the text is in secs. 3 and 4. 
All values are in MeV.}
\vspace{0.3cm}
\begin{tabular}{|c|r|r|r|r|}
\hline
             & \multicolumn{2}{|c|}{ISGMR} & \multicolumn{2}{|c|}{ISGDR} \\
             \cline{2-5}
             &  Exp.    & Theory    & Exp.    & Theory \\ 
\hline
$^{90}$Zr    & 17.9     & 19.1      & 26.2    & 27.1   \\
$^{116}$Sn   & 16.0     & 17.2      & 23.0    & 26.3   \\
$^{144}$Sm   & 15.3     & 16.2      & 24.2    & 25.4   \\
$^{208}$Pb   & 14.2     & 14.1      & 20.3    & 24.1   \\  
\hline
\end{tabular}
\label{table1}
\end{table}
\vspace{0.2cm}

\section*{Results for the isoscalar dipole resonance}

A peculiar feature of calculations of this giant resonance is the appearance of 
a spurious state in the calculated spectrum. 
When diagonalizing the RPA matrix on a 1$^-$ basis, we expect to see among 
%our results,  
all states 
the spurious state at zero energy corresponding to the center-of-mass motion 
and we expect as well that it exhausts the whole strength associated with the
operator $\sum_i r_i Y_{10}$. Due to a lack of complete self-consistency 
(some part of the residual interaction, like the two-body spin-orbit and Coulomb 
forces, are usually neglected in the RPA because their effect should be 
rather small) and to numerical inaccuracies, this is not the case. The 
spurious state 
%results 
comes out in practice 
at finite energy and its wave function does not
overlap completely with that of the exact center-of-mass motion: as a 
consequence, the remaining RPA eigenstates are not exactly orthogonal to the
true spurious state, and their spurious component must be projected out. 
This is not difficult if RPA is done in the discrete p-h space. 
%Anyway, 
In any case, 
it can be shown$^{16)}$ that this projection procedure is equivalent to
replacing the $\sum_i r_i^3 Y_{10}$ operator with $\sum_i (r_i^3-\eta r_i) 
Y_{10}$, $\eta$ being ${5\over 3}\langle r^2 \rangle$. Once this projection is 
done, we find that a substantial amount of $r^3 Y_{10}$ strength still remains 
in the 10 - 15 MeV region, in addition to the strength in the 20 - 25 MeV region. 
The data do not show any strength in the lower region, i.e., experimental centroid 
energies correspond to the energy region above 15 MeV. Therefore, to have a 
meaningful comparison with experiment we will refer from now on to theoretical 
centroid energies calculated in the interval 15 - 40 MeV. 

In Table 1 we show the ISGDR centroid energies obtained with the SGII 
force. Especially in the case of $^{208}$Pb and $^{116}$Sn, it can be noticed
that RPA calculations tend to overestimate the value of the centroid energy, 
the discrepancy being less severe in the other two cases. One may wonder 
if this is a special feature of SGII, although this force has been said to
behave rather well for the monopole case. Fig. 1 shows that this is not the
case: the centroid energies obtained in RPA with a number of different Skyrme
parametrizations are plotted as a function of their incompressibility $K$, and
it can be noticed that all forces systematically overestimate the experimental
values of the centroid energies. 
%We must stress at this point that 
Gogny 
interactions have also been used to study the ISGDR in this and other 
nuclei$^{17)}$, but they also predict too large centroid energies. 
%(...). 
The same can be said about relativistic models like relativistic RPA$^{18)}$ 
or time-dependent relativistic mean field$^{19)}$. 
We may conclude that the case of the 
ISGDR in $^{208}$Pb is a kind of exception among the giant resonances studied 
within the self-consistent HF-RPA approach, as usually one never finds such 
large discrepancies between theory and experiment ($\sim$ 4-5 MeV). 

We finally address the question whether the coupling with more complicated 
configurations, which has been seen to be responsible of a downward shift of the 
resonance energy, can diminish this discrepancy in the case of $^{208}$Pb. We 
have done for the ISGDR a calculation of the type described above in the monopole case. 
%also for the ISGDR. 
The resulting strength function, which includes continuum and 
2p-2h coupling, is depicted in Fig. 2. One can see that the total
width of the resonance is quite large, and the theoretical value of about 6 
MeV compares well with the experimental result which is about 7 MeV$^{20)}$. 
Although the resonance lineshape is accounted for by theory, the 
downward shift with respect to the RPA result is only about 1 MeV. The 
fundamental problem why the ISGDR energy in $^{208}$Pb cannot be reproduced by 
theoretical models still remains.
  
\section*{Conclusion}

In this paper, we have considered the isoscalar monopole and dipole resonances 
in a number of nuclei and we have tried to reproduce their properties by means of
HF-RPA, or QRPA on top of HF-BCS, or more sophisticated approach which takes care of
the continuum properly and of the coupling with 
%doorway states, that is, 
nuclear 
configurations which are more complicated than the simple 1p-1h. In general, we 
have found that the effect of pairing correlations is quite small as these 
resonances lie at high energy with respect to the pairing gap $\Delta$. 

Concerning the RPA results, the situation looks different in the case of monopole 
and dipole. In the former case, the Skyrme-type force SGII is able to reproduce 
well the centroid energy in $^{208}$Pb, and it slightly overestimates this energy 
%in the case of the 
in other medium-heavy nuclei which have been measured accurately in recent
experiments. This would allow us to extract a value of the nuclear incompressibility
around 215 MeV. In the case of the ISGDR, however, the same Skyrme force 
overpredicts this centroid energy in $^{208}$Pb by about 4 MeV. Other parametrizations
of Skyrme type cannot do better, and the problem is not solved if one turns
to Gogny interactions or to relativistic models. Therefore, although in other nuclei
this discrepancy between theory and experiment can be less than in the case of 
$^{208}$Pb, we can say that the ``squeezing mode'', which could be taken as a 
further probe of the nuclear incompressibility besides the well-known ``breathing'' 
monopole oscillation, is challenging us with a new problem.

Calculations beyond mean field do not change substantially our conclusions about the
centroid energies. However, 
we stress that these calculations are necessary for a proper
account of the giant resonances lineshape, and in fact in our case they have been
able to reproduce the total width of the ISGMR in $^{208}$Pb, and also of the 
ISGDR although its centroid is overestimated. 

We would like to thank Umesh Garg for stimulating discussions and 
for communicating experimental data prior to publication, and also Jean-Paul
Blaizot for useful discussions. 

\begin{figure}[h]%1
\caption{RPA centroid energies of ISGDR calculated in $^{208}$Pb with various 
Skyrme interactions, as a function of compression modulus $K$. }
\end{figure}

\begin{figure}[h]%2
\caption{Calculated ISGDR strength distribution of operator $(r^3 - \eta r) Y_{10}$ 
including damping effects of particle-hole-plus-phonon coupling (see text). The 
nucleus is $^{208}$Pb. }
\end{figure}

\section*{References}
\re
1) J.\ P.\ Blaizot: Phys.\ Rep.\ {\bf 64}, 171 (1980); J.\ M.\ Pearson: Phys.\ 
Lett.\ {\bf B 271}, 12 (1991); S.\ Shlomo and D.\ H.\ Youngblood: Phys.\ Rev.\ 
{\bf C 47}, 529 (1993).
\re
2) D.\ H.\ Youngblood, H.\ L.\ Clark and Y.- W.\ Lui: Phys.\ Rev.\ Lett.\ 
{\bf 82}, 691 (1999). 
\re
3) B.\ Davis {\it et al.}: Phys.\ Rev.\ Lett.\ {\bf 79}, 609 (1997).
\re
4) S.\ Stringari: Phys.\ Lett.\ {\bf B106}, 232 (1982).
\re
5) M.\ Beiner, H.\ Flocard, N.\ Van Giai and Ph.\ Quentin: Nucl.\ Phys.\ 
{\bf A238}, 29 (1975).
\re
6) N.\ Van Giai and H.\ Sagawa: Phys.\ Lett.\ {\bf B106}, 379 (1981).
\re
7) E.\ Chabanat, P.\ Bonche, P.\ Haensel, J.\ Meyer and R.\ Schaeffer: Nucl.\ 
Phys.\ {\bf A635}, 231 (1998).
\re
8) G.\ Col\`o, P.\ F.\ Bortignon, N.\ Van Giai, A.\ Bracco and R.\ A.\
Broglia: Phys.\ Lett.\ {\bf B276}, 279 (1992).
\re
9) A.\ Bohr and B.\ M.\ Mottelson, Nuclear Structure, vol. I, W.A. Benjamin
1969, Eqs. (2.92) and (2.93).
\re
10) N.\ Tajima, S.\ Takahara and N.\ Onishi: Nucl.\ Phys.\ {\bf A603}, 
23 (1996).
\re
11) E.\ Khan, N.\ Van Giai and G.\ Col\`o: to be published.
\re
12) G.\ Col\`o, N.\ Van Giai, P.\ F.\ Bortignon and R.\ A.\ Broglia:
Phys.\ Rev.\ {\bf C50}, 1496 (1994).
\re
13) J.\ P.\ Blaizot, J.\ F.\ Berger, J.\ Decharg\'e and M.\ Girod: 
Nucl.\ Phys.\ {\bf A591}, 435 (1995).
\re
14) O.\ Civitarese, A.\ G.\ Dumrauf, M.\ Reboiro, P.\ Ring and M.\ M.\ Sharma: 
Phys.\ Rev.\ {\bf C43}, 2622 (1991).
\re
15) I.\ Hamamoto, H.\ Sagawa and X.\ Z.\ Zhang: Phys.\ Rev.\ {\bf C56}, 3121 
(1997).
\re
16) N.\ Van Giai and H.\ Sagawa: Nucl.\ Phys.\ {\bf A371}, 1 (1981). 
\re
17) J.\ Decharg\'e and L.\ \v{S}ips: Nucl.\ Phys.\ {\bf A407}, 1 (1983). 
\re
18) N.\ Van Giai and Z.\ Y.\ Ma: these proceedings.
\re 
19) D.\ Vretenar {\it et al.}: these proceedings.  
\re
20)  U.\ Garg {\it et al.}: Proc. Topical Conference on Giant Resonances, 
Varenna 
( Nucl.\ Phys.\ {\bf A}, to be published; H.\ L.\ Clark {\it et al.}: ibid.\
;  U.\ Garg: private communication. 

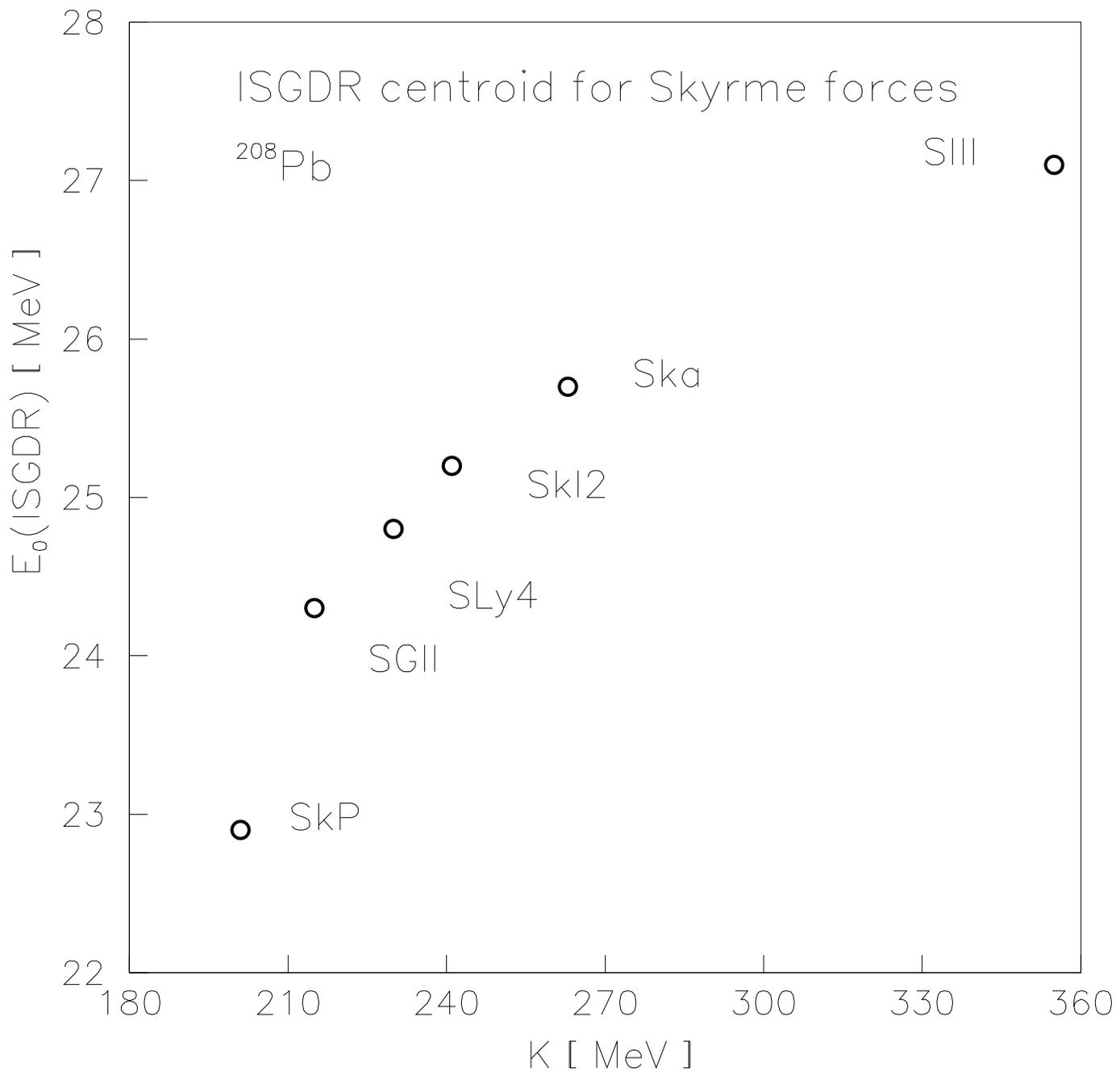
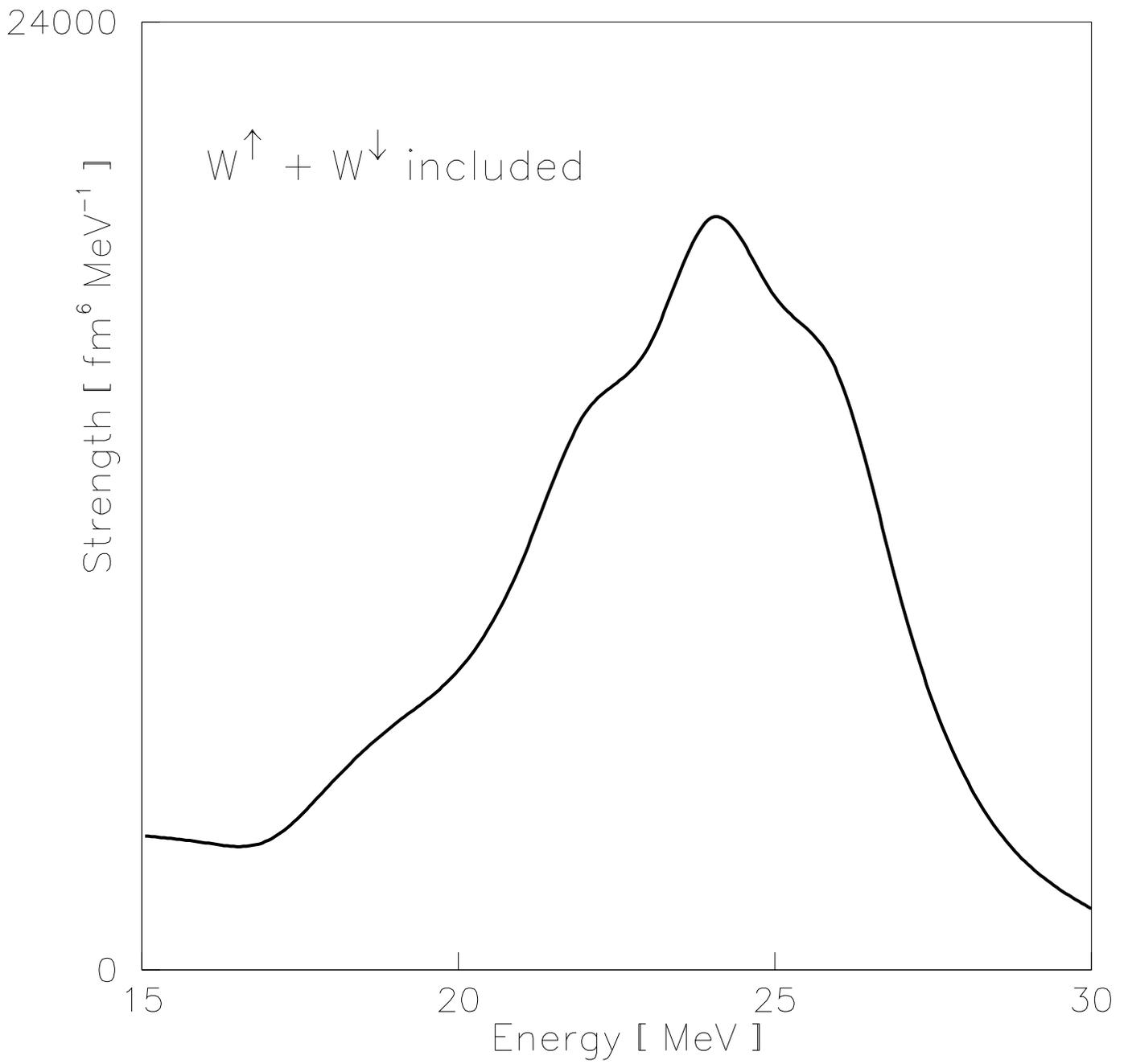
\end{document}